\documentclass{ifacconf}
\pdfoutput=1 
\usepackage[authoryear]{natbib}      
\usepackage{graphicx}
\usepackage{amsmath}
\usepackage{amssymb}
\usepackage{mathrsfs}
\usepackage{savesym}
\savesymbol{AND}
\usepackage{algorithmic}
\usepackage{algorithm}
\DeclareMathOperator{\supp}{supp}

\DeclareMathOperator{\tr}{Tr}
\DeclareMathOperator{\diag}{diag}
\begin{document}
\begin{frontmatter}
\title{Error analysis in Bayesian identification of non-linear state-space models}
\author[First]{Aditya Tulsyan},
\author[First]{Biao Huang},
\author[Second]{R. Bhushan Gopaluni},
\author[First]{J. Fraser Forbes}
\thanks[footnote]{This article has been published in: Tulsyan, A, B. Huang, R.B. Gopaluni and J.F. Forbes (2013). Bayesian identification of non-linear state-space models: Part II- Error Analysis.  In: \emph{Proceedings of the 10th IFAC International Symposium on Dynamics and Control of Process Systems}. Mumbai, India.}
\thanks[footnote]{This work was supported by the Natural Sciences and Engineering Research Council (NSERC), Canada.}

\address[First]{Department of Chemical and Materials Engineering, University of Alberta, Edmonton AB T6G-2G6, Canada (e-mail: \{tulsyan; biao.huang; fraser.forbes\}@ ualberta.ca)}
\address[Second]{Department of Chemical and Biological Engineering, University of
British Columbia, Vancouver BC V6T-1Z3, Canada (e-mail: gopaluni@chbe.ubc.ca)}
\begin{abstract}
In the last two decades, several methods based on sequential Monte Carlo (SMC) and Markov chain Monte Carlo (MCMC) have been proposed for Bayesian identification of stochastic non-linear state-space models (SSMs). It is well known that the performance of these simulation based identification methods depends on the numerical approximations used in their design. We propose the use of posterior Cram\'er-Rao lower bound (PCRLB) as a mean square error (MSE) bound. Using PCRLB, a systematic procedure is developed to analyse the estimates delivered by Bayesian identification methods in terms of bias, MSE, and efficiency. The efficacy and utility of the proposed approach is illustrated through a numerical example.
\end{abstract}
\end{frontmatter}
\section{Introduction}
\label{sec:S1}
Bayesian identification has a long history, dating at least as far back as \cite{P1981}. Despite this, it is not commonly used in practice, except for the linear, Gaussian SSM case; wherein, Kalman filter based Bayesian estimate is routinely employed (\cite{N2010}). This is due to the computational complexities associated with the computation of the posterior densities, their marginals, and associated functions, such as posterior mean and variance (\cite{J2005}). Recent developments in statistical methods, such as SMC and MCMC along with advances in computing technology have allowed researchers to use Bayesian methods in both on-line (\cite{T2013}, \cite{C2005}) and off-line (\cite{J2011}, \cite{G2001}) identification of SSMs. 

This paper is directed towards the class of Bayesian identification methods for parameter estimation in  stochastic SSMs. 
The notation used in this paper is introduced next.

\textit{Notation:} ${\mathbb{N}:=\{1,2,\dots\}}$; ${\mathbb{R}_+:=
[0,\infty)}$; $\mathbb{R}^{\rm s\times s}$ is the set of real-valued ${s\times s}$ matrices; ${\mathcal{S}^s\subset \mathbb{R}^{\rm s\times s}}$ is the space of symmetric matrices; $\mathcal{S}^s_{+}$ is the cone of symmetric positive semi-definite matrices in $\mathcal{S}^s$; and $\mathcal{S}_{++}^s$ is its interior. The partial order on $\mathcal{S}^s$ induced by $\mathcal{S}_{+}^s$ and $\mathcal{S}_{++}^s$ are denoted by $\succcurlyeq$ and $\succ$, respectively. For ${A\in\mathbb{R}^{\rm s\times s}}$, $\tr[A]$ denotes its trace. For a vector $y\in\mathbb{R}^p$, $\diag(y)\in\mathcal{S}^p$ is a diagonal matrix with $y\in\mathbb{R}^p$ as its entries. $\lvert\cdot\rvert$ is the absolute value. $\Delta^{y}_{x}\triangleq\nabla_{x}\nabla_{y}^T$ is Laplacian and $\nabla_{x}\triangleq\left[\frac{\partial{}}{\partial{x}}\right]$ is gradient. 
\section{Bayesian identification}
\label{sec:S2}
Let $\{X_t\}_{t\in\mathbb{N}}$ and $\{Y_t\}_{t\in\mathbb{N}}$ be ${\mathcal{X}} (\subseteq \mathbb{R}^{n})$ and ${\mathcal{Y}} (\subseteq \mathbb{R}^{m})$ valued stochastic processes defined on a measurable space $(\Omega, \mathcal{F})$. Let these stochastic processes depend on unknown parameter vector ${\theta} \in {\Theta}$, where ${\Theta}$ is an open subset of $\mathbb{R}^{q}$. The discrete-time state $\{X_t\}_{t\in\mathbb{N}}$ is an unobserved process, with initial density $p_{\theta}(x)$ and transition density $p_{\theta}(x'|x)$:
\begin{equation}
\label{eq:E1}
X_0\sim p_{\theta}(\cdot);~~X_{t+1}|(X_t=x_t)\sim p_\theta(\cdot|x_t,u_{t})~~(t\in\mathbb{N}).
\end{equation}
It is assumed that $\{Y_t\}_{t\in\mathbb{N}}$ is conditionally independent given $\{X_t\}_{t\in\mathbb{N}}$ and have a marginal density $p_{\theta}(y|x)$:
\begin{equation}
\label{eq:E2}
Y_{t}|(X_0,\dots,X_t=x_t,\dots,X_T)\sim p_\theta(\cdot|x_t)~~~\quad(t\in\mathbb{N}).  
\end{equation}
All the densities are with respect to suitable dominating measures, such as Lebesgue measure, which are denoted generically as $dx$ and $dy$. Although (\ref{eq:E1}) and (\ref{eq:E2}) represent a wide class of non-linear time-series models, the model form considered in this paper is given below
\begin{subequations}
\label{eq:E3}
\begin{align}
{X}_{t+1}&={f}_t({X}_{t},\theta_t,V_t);\label{eq:E3a}\\
\theta_{t+1}&=\theta_t;\label{eq:E3b}\\
{Y}_t&={g}_t({X}_{t}, \theta_t,W_t)\label{eq:E3c},
\end{align}
\end{subequations}
where ${\{\theta_t\}_{t\in\mathbb{N}}=\theta}$ is a vector of unknown parameters, and $\{V_t\}_{t\in\mathbb{N}}$ and $\{W_t\}_{t\in\mathbb{N}}$ are the state and measurement noise.
\begin{rem}
\label{R0}
To minimize use of notation, the input signal $\{u_t\}_{t\in\mathbb{N}}$ is not included in (\ref{eq:E3}); however, all the results that appear in this paper hold with signal $\{u_t\}_{t\in\mathbb{N}}$ included. \qed
\end{rem}
For a generic sequence $\{r_t\}_{t\in\mathbb{N}}$, let $r_{i:j}\triangleq\{r_i,r_{i+1},\dots,r_j\}$. In Bayesian identification, the problem of estimating the parameter vector ${\theta\in {\Theta}\subseteq\mathbb{R}^q}$ in (\ref{eq:E3}), given a measurement sequence ${\{Y_{1:t}=y_{1:t}\}_{t\in\mathbb{N}}}$ is formulated as a joint state and parameter estimation problem. This is done by ascribing a prior density ${\theta_0\sim p(\theta_0)}$, such that ${\theta\in\supp{p(\theta_0)}}$, and computing the density ${Z_t|(Y_{1:t}=y_{1:t})\sim p(\cdot|y_{1:t})}$, where ${Z_t\triangleq\{X_t;~\theta_t\}}$ is a ${\mathcal{Z} (\subseteq \mathbb{R}^{s=n+q})}$ valued extended Markov process with ${(Z_0=z_0)\sim p_{\theta_0}(x_0)p(\theta_0)}$, ${Z_t|(Z_{t-1}=z_{t-1})}$ ${\sim p_{\theta_{t-1}}(\cdot|x_{t-1})\delta_{\theta_{t-1}}(\cdot)}$. Note that a recursive method to compute $\{p(z_t|y_{1:t})\}_{t\in\mathbb{N}}$ is given by the optimal filtering equation. Having computed $\{p(z_t|y_{1:t})\}_{t\in\mathbb{N}}$, inference on $\{\theta_t\}_{t\in\mathbb{N}}$ then relies on the marginal density $\{p(\theta_t|y_{1:t})\}_{t\in\mathbb{N}}$. 

Although computing ${\theta_t|(Y_{1:t}=y_{1:t})\sim p(\cdot|y_{1:t})}$ appears  similar to computing ${X_t|(Y_{1:t}=y_{1:t})\sim p_{\theta}(\cdot|y_{1:t})}$ (under known parameter case) in the state estimation problem, calculating $\{p({\theta}_t | {y}_{1:t})\}_{t\in\mathbb{N}}$ for (\ref{eq:E3}) has proved to be a non-trivial problem (\cite{Minvielle2010}, \cite{Kantas2009}). No analytical solution to $\{p({\theta}_t | {y}_{1:t})\}_{t\in\mathbb{N}}$ is available, even for linear and Gaussian SSM, or when $\mathcal{X}$ is a finite set (\cite{Kantas2009}). There are several simulation and numerical methods (e.g., SMC, MCMC, Kalman based filters), which allow for recursive approximation of $\{p(\theta_t|y_{1:t})\}_{t\in\mathbb{N}}$. Although tractable, the quality of these identification methods depends on the underlying numerical and statistical approximations used in their design.

Despite the widespread interest in developing advanced simulation and numerical methods for Bayesian identification of (\ref{eq:E3}), there have been no elaborate study on the quality of these methods. With this background, this paper proposes the use of PCRLB as an error bound. Using PCRLB, a systematic approach to assess the quality of a Bayesian identification method, in terms of bias, MSE, and efficiency is developed. Initial results reported by the authors in \cite{Tulsyan2013b} use PCRLB for assessment of state (but not parameter) estimation algorithms. The focus of this paper is to extend the results in \cite{Tulsyan2013b} to the Bayesian parameter estimation algorithms.
\section{PCRLB as an Error bound}
\label{sec:S3}
The conventional Cram\'er-Rao lower bound (CRLB) provides a theoretical lower bound on the MSE of any maximum-likelihood (ML) based unbiased parameter estimator. An analogous extension of the CRLB to the Bayesian estimators was derived by \cite{V1968}, and is commonly referred to as the PCRLB inequality. The PCRLB, derived recently by \cite{T1998} for (\ref{eq:E3}), provides a lower bound on the MSE associated with the joint estimation of the states and parameters from $\{p(z_t|u_{1:t}, y_{1:t})\}_{t\in\mathbb{N}}$, and is given in the next lemma.
\begin{lem}
\label{L4}
Let ${\{Y_{1:t}=y_{1:t}\}_{t\in\mathbb{N}}}$ be an output sequence generated from (\ref{eq:E3}), then the MSE associated with the estimation of $\{Z_t\}_{t\in\mathbb{N}}$ from $\{p(z_t|y_{1:t})\}_{t\in\mathbb{N}}$ is bounded by 
\begin{equation}
\label{eq:E4}
P^z_{{t|t}}\triangleq\mathbb{E}_{p(z_{t}, y_{1:t})}[(Z_t-{Z}_{t|t})(Z_t-{Z}_{t|t})^T]\succcurlyeq [J_t^z]^{-1},
\end{equation}
where: ${{Z}_{t|t}:=\mathbb{R}^{tm}\rightarrow \mathbb{R}^s}$ is a point estimate of ${\{Z_t\}_{t\in\mathbb{N}}}$; ${P^z_{t|t}\triangleq\left[
  \begin{array}{cc}
    P^{x}_{t|t} & P^{x\theta}_{t|t} \\
    (P^{x\theta}_{t|t})^T & P^{\theta}_{t|t} \\
  \end{array}
\right]\in\mathcal{S}^{s}_{++}}$, ${J^z_{t}\triangleq\left[
  \begin{array}{cc}
    J_{t}^{x} & J_{t}^{x\theta} \\
    (J_{t}^{x\theta})^T & J_{t}^{\theta} \\
  \end{array}
\right]\in\mathcal{S}^{s}_{++}}$, ${[J_t^z]^{-1}\triangleq\left[
  \begin{array}{cc}
    L_{t}^x & L_{t}^{x\theta} \\
    (L_{t}^{x\theta})^T & L_{t}^\theta \\
  \end{array}
\right]\in\mathcal{S}^{s}_{++}}$ are  the MSE, posterior information matrix (PIM), and PCRLB, respectively.
\end{lem}
\begin{pf}
See \cite{T1998} for proof. ~~~~~~~~~~~~~~~\qed
\end{pf}
A recursive approach to compute ${J_t^z\in\mathcal{S}^{s}_{++}}$   was derived by \cite{T1998}, and is given next. But first, we give the assumptions on the model considered in (\ref{eq:E3}).
\begin{assum}
\label{A1}
$\{V_t\}_{t\in\mathbb{N}}$ and $\{W_t\}_{t\in\mathbb{N}}$ are mutually independent sequences of independent random variables known a priori in their distribution classes (e.g., Gaussian) and parametrized by a known and finite number of moments.
\end{assum}
\begin{assum}
\label{A2}
${f_t:=\mathcal{X}\times\Theta\times\mathbb{R}^m\rightarrow\mathbb{R}^n}$ and $g_t:=\mathcal{X}\times$ ${\Theta\times\mathbb{R}^m\rightarrow\mathbb{R}^m}$ are non-linear functions, such that in the open set $\mathcal{X}$ and $\Theta$, $\{f_t;g_t\}$ is $\mathcal{C}^k({\mathcal{X}})$ and $\mathcal{C}^k(\Theta)$, and in $\mathbb{R}^n$ and $\mathbb{R}^m$, $f_t$ is $\mathcal{C}^{k-1}({{\mathbb{R}^n}})$, and $g_t$ is $\mathcal{C}^{k-1}({{\mathbb{R}^m}})$, where $k\geq2$. 
\end{assum}
\begin{assum}
\label{A3}
For any random sample $(x_{t+1},x_t,\theta_t,v_t)$ ${\in\mathcal{X}\times\mathcal{X}\times\Theta\times\mathbb{R}^n}$ and $(y_t,x_t,\theta_t,w_t)\in\mathcal{Y}\times\mathcal{X}\times\Theta\times\mathbb{R}^m$ satisfying (\ref{eq:E3}), $\nabla_{v_t}f^T_t(x_t,\theta_t,v_t)$ and $\nabla_{w_t}g^T_t(x_t,\theta_t,w_t)$ have rank $n$ and $m$, respectively, such that using implicit function theorem, $p_\theta(x_{t+1}|x_t)=p(V_t=\tilde{f}_t(x_t,\theta_t,x_{t+1}))$ and $p_{\theta}(y_{t}|x_t)=p(W_t=\tilde{g}_t(x_t,\theta_t,y_{t}))$ are defined. 
\end{assum}
\begin{lem}
\label{L5}
A recursive approach to compute ${J_t^z\in\mathcal{S}^{s}_{++}}$ for (\ref{eq:E3}) under Assumptions \ref{A1} through \ref{A3} is given as follows
\begin{subequations}
\label{eq:E5}
\begin{align}
J^{x}_{t+1}&=H_t^{33}-(H_t^{13})^T[J_t^{x}+H_t^{11}]^{-1}H_t^{13};\label{eq:E5a}\\
J^{x\theta}_{t+1}&=(H_t^{23})^T-(H_t^{13})^T[J_t^{x}+H_t^{11}]^{-1}(J_t^{x\theta}+H_t^{12});\label{eq:E5b}\\
J^{\theta}_{t+1}&=J_t^{\theta}+H_t^{22}-(J_t^{x\theta}+H_t^{12})^T[J_t^{x}+H_t^{11}]^{-1}
\nonumber\\
&\times(J_t^{x\theta}+H_t^{12}),\label{eq:E5d}
\end{align}
\end{subequations}
where:
\begin{subequations}
\label{eq:E6}
\begin{align}
H_t^{11}&=\mathbb{E}_{p(x_{0:t+1},\theta_t,y_{1:t+1})}[-\Delta_{X_t}^{X_t}\log{p_t}];\label{eq:E6a}\\
H_t^{12}&=\mathbb{E}_{p(x_{0:t+1},\theta_t,y_{1:t+1})}[-\Delta_{X_t}^{\theta_t}\log p_t];\label{eq:E6b}\\
H_t^{13}&=\mathbb{E}_{p(x_{0:t+1},\theta_t,y_{1:t+1})}[-\Delta_{X_t}^{X_{t+1}}\log p_t];\label{eq:E6c}\\
H_t^{22}&=\mathbb{E}_{p(x_{0:t+1},\theta_t,y_{1:t+1})}[-\Delta_{\theta_t}^{\theta_t}\log{p}_{t}];\label{eq:E6d}\\
H_t^{23}&=\mathbb{E}_{p(x_{0:t+1},\theta_t,y_{1:t+1})}[-\Delta_{\theta_t}^{X_{t+1}}\log{p}_{t}];\label{eq:E6e}\\
H_t^{33}&=\mathbb{E}_{p(x_{0:t+1},\theta_t,y_{1:t+1})}[-\Delta_{X_{t+1}}^{X_{t+1}}\log{p}_{t}];\label{eq:E6f}
\end{align}
\end{subequations}
and: ${{p}_{t}=p(X_{t+1}|Z_t)p(Y_{t+1}|\theta_t,X_{t+1})}$; and the PIM at $t=0$ can be computed using $J_0=\mathbb{E}_{p(z_0)}[-\Delta_{Z_0}^{Z_0}\log{p}(Z_0)]$.
\end{lem}
\begin{pf}
See \cite{T1998} for proof.\qquad\qquad\qquad\qed
\end{pf}
Since the focus here is on $\{\theta\}_{t\in\mathbb{N}}$ alone, a lower bound on the MSE associated with the estimation of $\{\theta\}_{t\in\mathbb{N}}$ is of interest to us. Using Lemmas \ref{L4} and \ref{L5}, a bound on the MSE for parameter estimates can be derived, as given next. 
\begin{cor}
\label{C6}
Let ${P^z_{{t|t}}\in\mathcal{S}^{s}_{++}}$ and ${J_t^z\in\mathcal{S}^{s}_{++}}$ be such that they satisfy (\ref{eq:E4}), then the MSE associated with the estimation of $\{\theta_t\}_{t\in\mathbb{N}}$ from $\{p(\theta_t|y_{1:t})\}_{t\in\mathbb{N}}$, is bounded by
\begin{equation}
\label{eq:E7}
P^\theta_{{t|t}}=\mathbb{E}_{p(\theta_{t}, y_{1:t})}[(\theta_t-{\theta}_{t|t})(\theta_t-{\theta}_{t|t})^T]\succcurlyeq L_t^\theta,
\end{equation}
where ${\theta_{t|t}:=\mathbb{R}^{tm}\rightarrow\mathbb{R}^q}$ is the parameter estimate delivered by a Bayesian identification algorithm, and $L_t^\theta\in\mathcal{S}^{q}_{++}$ is the lower right matrix of $[J_t^z]^{-1}\in\mathcal{S}^{s}_{++}$ in Lemma \ref{L4}.
\end{cor}
\begin{pf}
The proof is based on the fact that the inequality in Lemma \ref{L4} guarantees that $P^z_{{t|t}}-[J_t^z]^{-1}\in\mathcal{S}^{s}_{+}$. ~~~~~~~~~~~\qed
\end{pf}
A recursive approach to compute $L_t^\theta\in\mathcal{S}^{q}_{++}$ is given next.
\begin{thm}
\label{T7}
Let ${J_t^z\in\mathcal{S}^{s}_{++}}$ be the PIM for ${\{Z_t\}_{t\in\mathbb{N}}}$, and $L_t^\theta\in\mathcal{S}^{q}_{++}$ be the lower bound on the MSE associated with the estimation of $\{\theta_t\}_{t\in\mathbb{N}}$ in (\ref{eq:E3}), then given $J^z_t\in\mathcal{S}^{s}_{++}$, $L_t^\theta\in\mathcal{S}^{q}_{++}$ at $t\in\mathbb{N}$ can be recursively computed as follows
\begin{equation}
\label{eq:E8}
L_t^{\theta}=[J^\theta_t-(J_t^{x\theta})^T(J_t^{x})^{-1}J_t^{x\theta}]^{-1},
\end{equation}
where $J^\theta_t$, $J_t^{x\theta}$ and $J_t^{x}$ are the PIMs given in {Lemma \ref{L4}}.
\end{thm}
\begin{pf}
The proof is based on the matrix inversion lemma (see \cite{bapat1997}).\qquad~~~~~~~~\qed
\end{pf}
\begin{rem}
\label{R8}
Theorem \ref{T7} shows that for (\ref{eq:E3}), $L_t^{\theta}$ is not only a function of the PIM for $\{\theta_t\}_{t\in\mathbb{N}}$, i.e., $J_t^\theta$, but it also depends on the PIMs for $\{X_t\}_{t\in\mathbb{N}}$, i.e., $J_t^{x\theta}$ and $J_t^{x}$.~~~~~~\qed
\end{rem}
\begin{rem}
\label{R9}
Integral in (\ref{eq:E6}) with respect to ${p(x_{0:t},\theta_{t-1},y_{1:t})}$ makes $L_t^{\theta}$ in (\ref{eq:E8}) independent of any random sample from $\mathcal{X}^{t+1}$, $\Theta$, and $\mathcal{Y}^{t}$. $L_t^{\theta}$ in fact only depends on: the process dynamics in (\ref{eq:E3}); noise characteristics of $V_t\sim p(v_t)$ and $W_t\sim p(w_t)$; and the choice of  ${Z_0\sim p(z_0)}$. This makes $L_t^{\theta}$ a system property, independent of any Bayesian identification method or any specific realization from $\mathcal{X}$, ${\Theta}$ or $\mathcal{Y}$. This motivates the use of PCRLB as a benchmark for error analysis of Bayesian identification algorithms. ~~~~~~~~~~~\qed 
\end{rem}
Finally, using the inequality in (\ref{eq:E7}), the MSE associated with the parameter estimates obtained with any Bayesian identification method can be compared against the theoretical lower bound. Our approach to systematically compare and analyse the MSE and PCRLB is discussed next.
\section{PCRLB Inequality based Error analysis}
\label{sec:S4}
A common approach to compute ${\theta}_{t|t}\in\mathbb{R}^q$, is to minimize ${\tr[P^\theta_{t|t}]\in\mathbb{R}_+}$. This ensures that ${\tr[P^\theta_{t|t}-L_t^\theta]\geq 0}$ is minimized. The optimal estimate that minimizes ${\tr[P^\theta_{t|t}]\in\mathbb{R}_+}$ is referred to as the minimum MSE (MMSE) estimate, and is the conditional mean of ${\theta_t|(Y_{1:t}=y_{1:t})\sim p(\cdot|y_{1:t})}$, i.e., ${\theta_{t|t}={\theta}^\star_{t|t}\triangleq\mathbb{E}_{p(\theta_t|y_{1:t})}[\theta_t]}$ (see \cite{V1968} for derivation).
\begin{rem}
\label{R10}
Bayesian identification methods only approximate the true density $\{p(\theta_t|y_{1:t})\}_{t\in\mathbb{N}}$, thus in practice, the estimate delivered by identification methods may not be an MMSE estimate, i.e., ${{\theta}_{t|t}\triangleq\mathbb{E}_{\tilde{p}(\theta_t|y_{1:t})}[\theta_t]\neq \theta^\star_{t|t}}$ almost surely, where ${\theta}_{t|t}$ is the mean of ${\theta_t|(Y_{1:t}=y_{1:t})\sim\tilde{p}(\cdot|y_{1:t})}$ and $\{\tilde{p}(\theta_t|y_{1:t})\}_{t\in\mathbb{N}}$ is the approximate posterior.~~~~~~~~~~~~~~\qed
\end{rem}
The second-order error associated with $\{\theta_{t|t}\}_{t\in\mathbb{N}}$ is completely characterized by its MSE. A thorough assessment of any identification algorithm or that of its estimates requires clear understanding of the MSE. The next theorem shows decomposition of the MSE into its sources of errors.
\begin{thm}
\label{T11}
Let ${\theta}^\star_{t|t}\in\mathbb{R}^q$ and $V^\star_{t|t}\in\mathcal{S}^q_{++}$ be the mean and covariance of ${\theta_t|(Y_{1:t}=y_{1:t})\sim {p}(\cdot|y_{1:t})}$ and ${\theta}_{t|t}\in\mathbb{R}^q$ be the mean of $\theta_t|(Y_{1:t}=y_{1:t})\sim \tilde{p}(\cdot|y_{1:t})$ computed by a Bayesian identification method, then for ${\theta}_{t|t}\neq \theta^\star_{t|t}$ almost surely, $P^\theta_{t|t}$ at $t\in\mathbb{N}$ can be decomposed and written as
\begin{align}
\label{eq:E9}
P^\theta_{t|t}=\mathbb{E}_{p(y_{1:t})}[{V}^\star_{t|t}]+ \mathbb{E}_{p(y_{1:t})}[B^\star_{{t|t}}[B^\star_{{t|t}}]^T],
\end{align}
where ${B^\star_{{t|t}}\triangleq[\theta^\star_{t|t}-{\theta}_{t|t}]\in\mathbb{R}^q}$ is the conditional bias in estimating the true conditional mean ${\theta}^\star_{t|t}\in\mathbb{R}^q$ at $t\in\mathbb{N}$.
\end{thm}
\begin{pf}
The proof is adapted from (\cite{Tulsyan2013b}). From the definition of  expectation, MSE in (\ref{eq:E7}) can be written as $P^\theta_{t|t}=\mathbb{E}_{p(y_{1:t})}\mathbb{E}_{p(\theta_{t}|y_{1:t})}[(\theta_t-{\theta}_{t|t})(\theta_t-{\theta}_{t|t})^T]$, where we have used $p(\theta_t,y_{1:t})=p(y_{1:t})p(\theta_t|y_{1:t})$. Adding and subtracting ${\theta}^\star_{t|t}$ in $P^\theta_{t|t}$, followed by several algebraic manipulations yield
$P^\theta_{t|t}=\mathbb{E}_{p(y_{1:t})}\mathbb{E}_{p(\theta_{t}|y_{1:t})}[F^\star_{{t|t}}+G^\star_{{t|t}}+[G^\star_{{t|t}}]^T+B^\star_{{t|t}}[B^\star_{{t|t}}]^T]$, where $F^\star_{{t|t}}=[\theta_t-{\theta}^\star_{t|t}][\theta_t-{\theta}^\star_{t|t}]^T$; $G^\star_{{t|t}}=[\theta_t-{\theta}^\star_{t|t}][\theta_t-{\theta}_{t|t}]^T$. Now $\mathbb{E}_{p(\theta_{t}|y_{1:t})}[F^\star_{{t|t}}]=V^\star_{t|t}$; ${\mathbb{E}_{p(\theta_{t}|y_{1:t})}[G^\star_{{t|t}}]=0}$, since ${\mathbb{E}_{p(\theta_{t}|y_{1:t})}[\theta_t-{\theta}^\star_{t|t}]=0}$; and ${\mathbb{E}_{p(\theta_{t}|y_{1:t})}[B^\star_{{t|t}}][B^\star_{{t|t}}]^T=[B^\star_{{t|t}}][B^\star_{{t|t}}]^T}$, since ${[B^\star_{{t|t}}][B^\star_{{t|t}}]^T}$ is independent of ${\theta_t|(Y_{1:t}=y_{1:t})}$. Substituting the results into $P^\theta_{t|t}$  yields (\ref{eq:E9}), which completes the proof.~~~~~~~~~~~~~~~~~~\qed
\end{pf}
Note that Theorem \ref{T11} is the Bayesian equivalent of the classical MSE decomposition results available for the likelihood based estimators.
Using Theorem \ref{T11}, bias in the Bayesian parameter estimates $\{\theta_{t|t}\}_{t\in\mathbb{N}}$ is defined next.
\begin{defn}
\label{D2}
${\{\theta_{t|t}\}_{t\in\mathbb{N}}\in\mathbb{R}^q}$ is unconditionally unbiased if ${{\mathbb{E}_{p(y_{1:t})}[B^\star_{{t|t}}]=0}}$, and conditionally unbiased if ${{B^\star_{{t|t}}=0}}$ almost surely. The estimate which is both conditionally and unconditionally unbiased is an unbiased estimate.~\qed
\end{defn}
Bias in $\theta_{t|t}\in\mathbb{R}^q$ can be similarly defined as Definition \ref{D2}. The condition under which an identification method delivers unbiased parameter estimate is discussed next.
\begin{thm}
\label{T13}
Let ${{\theta}_{t|t}\in\mathbb{R}^q}$ be the estimate of ${{\theta}^\star_{t|t}\in\mathbb{R}^q}$, as computed by an identification method, where ${{\theta}^\star_{t|t}\in\mathbb{R}^q}$ is the mean of ${\theta_t|(Y_{1:t}=y_{1:t})\sim {p}(\cdot|y_{1:t})}$), and let ${B^\star_{{t|t}}\in\mathbb{R}^q}$ be the corresponding conditional bias,  
then $B^\star_{{t|t}}=0$ almost surely is only a sufficient condition for $\mathbb{E}_{p(y_{1:t})}[B^\star_{{t|t}}]=0$, but sufficient and necessary for $\mathbb{E}_{p(y_{1:t})}[B^\star_{{t|t}}[B^\star_{{t|t}}]^T]=0$.
\end{thm}
\begin{pf}
See \cite{Bill1995} for proof.~~~~~~~~~~~~~~~~~~~~~~~~~~~~~~~\qed
\end{pf}
\begin{rem}
\label{R14}
Theorem \ref{T13} shows that if the parameter estimate ${{\theta}_{t|t}\in\mathbb{R}^q}$ is unconditionally unbiased, it does not imply it is unbiased as well, but if it is conditionally unbiased, it implies ${{\theta}_{t|t}\in\mathbb{R}^q}$ is unbiased as well.~~~~~~~~~~~\qed 
\end{rem}

The MSE for an unbiased estimate ${{\theta}_{t|t}\in\mathbb{R}^q}$ is given next.
\begin{cor}
\label{C15}
Let ${{\theta}_{t|t}\in \mathbb{R}^q}$ be the estimate of the mean of ${\theta_t|(Y_{1:t}=y_{1:t})\sim {p}(\cdot|y_{1:t})}$ computed by a Bayesian identification method, such that ${B^\star_{{t|t}}=0}$ almost surely, then the MSE associated with ${{\theta}_{t|t}\in \mathbb{R}^q}$ is ${P^\theta_{t|t}=\mathbb{E}_{p(y_{1:t})}[V^\star_{t|t}]}$.~~~~~~\qed
\end{cor}
\begin{defn}
\label{D3}
An identification method delivering an estimate ${{\theta}_{t|t}\in \mathbb{R}^q}$ is efficient at $t\in\mathbb{N}$ if ${\tr[P^\theta_{t|t}-L_t^\theta]=0}$.\qed 
\end{defn}
\begin{thm}
\label{Th7}
Let ${{\theta}_{t|t}\in\mathbb{R}^q}$ be the estimate of ${{\theta}^\star_{t|t}\in\mathbb{R}^q}$, as computed by an identification method, and let $B^\star_{{t|t}}\in\mathbb{R}^s$ be the conditional bias in estimating ${{\theta}^\star_{t|t}\in\mathbb{R}^q}$, then ${B^\star_{{t|t}}=0}$ almost surely is both necessary and sufficient condition for the identification method to be efficient.
\end{thm}
\begin{pf}
For ${{\theta}_{t|t}\in\mathbb{R}^q}$ satisfying ${B^\star_{{t|t}}=0}$ almost surely, the MSE is given by ${P^\theta_{t|t}=\mathbb{E}_{p(Y_{1:t})}[V^\star_{t|t}]}$ (see Corollary \ref{C15}). Since $P^\theta_{t|t}$ only depends on $V^\star_{t|t}$, which is the covariance of ${\theta_t|(Y_{1:t}=y_{1:t})\sim p(\cdot|Y_{1:t})}$, $P^\theta_{t|t}$ cannot be reduced any further i.e., ${P^\theta_{t|t}=L_t^\theta}$. Thus from Definition \ref{D3} the identification method delivering ${{\theta}_{t|t}\in\mathbb{R}^q}$ is efficient at $t\in\mathbb{N}$.~~\qed
\end{pf}
Finally, the procedure to systematically assess the quality of the parameter estimates obtained with any Bayesian identification method is summarized in the next theorem.
\begin{thm}
\label{T17}
Let ${L^\theta_t\in\mathcal{S}^q_{++}}$ be the PCRLB on (\ref{eq:E3}), and let ${{\theta}^\star_{t|t}\in\mathbb{R}^q}$ and ${V^\star_{t|t}\in\mathcal{S}^q_{++}}$ be the mean and covariance of ${{\theta_t|(Y_{1:t}=y_{1:t})\sim {p}(\cdot|y_{1:t})}}$. Now if ${{\theta}_{t|t}\in\mathbb{R}^q}$ is an estimate of ${{\theta}^\star_{t|t}\in\mathbb{R}^q}$, as computed by an identification method, such that ${B^\star_{{t|t}}\in\mathbb{R}^q}$ is the conditional bias in estimating ${{\theta}^\star_{t|t}\in\mathbb{R}^q}$,  then for ${P^\theta_{t|t}\in\mathcal{S}^q_{++}}$ as the associated MSE, the quality of the estimate ${{\theta}_{t|t}\in\mathbb{R}^q}$ can be assessed as follows:\\
(a) If $B^\star_{{t|t}}=0$ almost surely, then (\ref{eq:E7}) is given by 
\begin{equation}
\label{eq:E10}
P^\theta_{t|t}=\mathbb{E}_{p(y_{1:t})}[V^\star_{t|t}]=L^\theta_t,
\end{equation}
which implies the identification method is efficient, and the corresponding estimate ${{\theta}_{t|t}\in\mathbb{R}^q}$ is unbiased and MMSE.\\
(b) If  ${B^\star_{{t|t}}\neq0}$ almost surely, then (\ref{eq:E7}) is given by 
\begin{equation}
\label{eq:E11}
P^\theta_{t|t}=\mathbb{E}_{p(y_{1:t})}[V^\star_{{t|t}}]+\mathbb{E}_{p(y_{1:t})}[B^\star_{{t|t}} [B^{\star}_{{t|t}}]']\succ L_t^{\theta},
\end{equation}
which implies the identification method is not efficient, and the  estimate ${{\theta}_{t|t}\in\mathbb{R}^q}$ is biased (only conditionally biased if $\mathbb{E}_{p(y_{1:t})}[B^\star_{{t|t}}]=0$) and not an MMSE estimate.
\end{thm}
\begin{pf}
The proof is based on the collective developments of Section \ref{sec:S4}, and is omitted here for the sake of brevity.~\qed
\end{pf}
The PCRLB inequality based error analysis tool developed in this section allows for assessment of parameter estimates obtained with Bayesian identification methods; however, obtaining a closed form solution to (\ref{eq:E7}) is non-trivial for (\ref{eq:E3}). Use of numerical methods is discussed next.
\section{Numerical methods}
\label{sec:S5}
It is well known that computing the MSE and PCRLB in (\ref{eq:E7}) in closed form is non-trivial for  the model considered in (\ref{eq:E3}) (see \cite{T1998},  \cite{NB2001}). This is because of the complex, high-dimensional integrals in the MSE with respect to $p(\theta_t,y_{1:t})$ (see (\ref{eq:E7})) and in the PCRLB with respect to $p(x_{0:t},\theta_{t-1},y_{1:t})$ (see (\ref{eq:E6a}) through (\ref{eq:E6f})), which do not admit any analytical solution. 

To address this issue, we use Monte Carlo (MC) sampling to numerically compute the MSE and PCRLB in (\ref{eq:E7}). For the sake of brevity, the procedure for MC approximation of the PCRLB is not provided here, but can be found in \cite{Tulsyan2013D}; however, for completeness, we provide an example for computation of MC based MSE.
\begin{exmp}
\label{Ex5}
Simulating samples ${\{(\theta_{t}=\theta^{j}_{t}, Y_{1:t}=y^j_{1:t})\}_{j=1}^{M}}$ ${\sim p(\theta_{t},y_{1:t})}$, $M$ times using (\ref{eq:E3}), starting at $M$ i.i.d.~initial draws from ${\{\theta_0\}_{i=1}^M\sim p(\theta_0)}$ and computing the estimates $\{{\theta}^j_{t|t}\}_{j=1}^M$, the MSE ${P}^\theta_{t|t}$ at $t\in\mathbb{N}$ can be approximated as
\begin{align}
\label{eq:E12}
\tilde{P}^\theta_{t|t}=\frac{1}{M}\sum_{j=1}^{M}(\theta^{j}_t-{\theta}^j_{t|t})(\theta^{j}_t-{\theta}^j_{t|t})^T,
\end{align}
where ${\tilde{P}^\theta_{t|t}\in\mathcal{S}_{++}^q}$ is an $M$-sample MC estimate of ${P}^\theta_{t|t}$. \qed
\end{exmp}  
Since (\ref{eq:E12}) is based on perfect sampling, using strong law of large numbers ${\tilde{P}_{t|t}^{\theta}\xrightarrow[]{a.s.}{P}_{t|t}^{\theta}}$ as ${M\rightarrow +\infty}$, where ${\xrightarrow{a.s.}}$ denotes almost sure convergence (see \cite{DM2004}). Note that $\tilde{L}_t^{\theta}$, which is an $M$-sample MC estimate of ${L}_t^{\theta}$ can also be similarly approximated using MC sampling. Details are omitted here, but can be found in \cite{Tulsyan2013D}. Despite the convergence proof, there are practical issues with the use of numerical methods, as given next.
\begin{rem}
\label{R19}
With ${M<+\infty}$, the MC estimate of the MSE and PCRLB may not necessarily satisfy the positive semi definite condition $\tilde{P}^\theta_{t|t}-\tilde{L}_t^\theta\succcurlyeq 0$ for all $t\in\mathbb{N}$.~~~~~~~~~~~~~\qed
\end{rem}
\begin{rem}
\label{R13}
Since $M<+\infty$, the conditions in Theorem \ref{T17} are relaxed to $\lvert B^\star_{{t|t}}\rvert\leq\epsilon$ and $\lvert \mathbb{E}_{p(Y_{1:t})}[B^\star_{{t|t}}]\rvert\leq\alpha$, and $\epsilon\in\mathbb{R}^{q}_{+}$ and $\alpha\in\mathbb{R}^q_+$ are pre-defined tolerance levels set based on $M$ and the required degree of accuracy.~~~~~~~~~~~~\qed
\end{rem}
\begin{rem}
\label{R14}
An identification method satisfying $\lvert B^\star_{{t|t}}\rvert\leq\epsilon$ is $\epsilon$-efficient at $t\in\mathbb{N}$ and the corresponding estimate is $\epsilon$-unbiased and $\epsilon$-MMSE (see Theorem \ref{T17}(a)). Similarly, if the estimate only satisfies $\lvert \mathbb{E}_{p(y_{1:t})}[B^\star_{{t|t}}]\rvert\leq\alpha$, then it is $\alpha$-unconditionally unbiased (see Theorem \ref{T17}(b)).~~~~~~~~~~~~\qed
\end{rem}
\section{Final algorithm}
A systematic approach to assess the quality of a Bayesian identification method,  proposed in Sections \ref{sec:S3} through \ref{sec:S5} is formally outlined in Algorithm \ref{alg:Algo1}.
\newcounter{ALC@tempcntr}
\newcommand{\LCOMMENT}[1]{%
    \setcounter{ALC@tempcntr}{\arabic{ALC@rem}}
    \setcounter{ALC@rem}{1}
    \item 
    \setcounter{ALC@rem}{\arabic{ALC@tempcntr}}
}%

\begin{algorithm}[!h]
  \caption{Analysis of Bayesian identification methods}
  \label{alg:Algo1}
  \begin{algorithmic}[1]
  \LCOMMENT ~\begin{center} \textit{Module 1: Computing the lower bound}  \end{center}
 \LCOMMENT ~ \textbf{Input:} Given (\ref{eq:E3}), define ${Z_t=\{X_t,\theta_t\}}$ and assume a prior density on $\{Z_t\}_{t\in\mathbb{N}}$, such that ${(Z_0=z_0)\sim p(z_0)}$
 \LCOMMENT ~ \textbf{Output:} Lower bound on the system in (\ref{eq:E3})
  \STATE Generate $M$ i.i.d.~samples from the assumed prior density $Z_0\sim p(\cdot)$, such that ${\{(Z_0=z_0^i)\}_{i=1}^M\sim p(z^i_0)}$
    \FOR{$t=1$ to $T$}
    \STATE Generate $M$ random samples from the states ${\{X_t=x_t^i|(Z_{t-1}=z_{t-1})\}_{i=1}^M\sim p(x_t^i|z_{t-1})}$ using (\ref{eq:E3a})
    \STATE Generate $M$ random samples from the parameters ${\{\theta_t=\theta_t^i|(Z_{t-1}=z_{t-1})\}_{i=1}^M\sim p(\theta_t^i)}$ using (\ref{eq:E3b}).  Note that in this step $\theta^i_t=\theta_0^i$ for all $1\leq i\leq M$ (see (\ref{eq:E3b}))
    \STATE Generate $M$ random samples from the measurements ${\{Y_t=y_t^i|(Z_{t}=z^i_{t})\}_{i=1}^M\sim p(y_t^i|z^i_t)}$ using (\ref{eq:E3c})
    \STATE Compute an $M$-sample MC estimate of $\tilde{J}_t^z$
    \STATE Compute an $M$-sample MC estimate of $\tilde{L}_t^\theta$
    \ENDFOR
    \LCOMMENT ~\begin{center} \textit{Module 2: Computing the estimates}  \end{center}
    \LCOMMENT ~ \textbf{Input:} Measurement sequences from Module 1, denoted as ${\{(Y_{1:T}=y^i_{1:T})\}_{i=1}^M}$ and a Bayesian identification method, which can compute $\{p(\theta_t|y_{1:t})\}_{t\in\mathbb{N}}$ (e.g., SMC, MCMC, EKF, and UKF)
    \LCOMMENT ~ \textbf{Output:} Parameter estimates
    \FOR{$i=1$ to $M$}
    \FOR{$t=1$ to $T$}
    \STATE Compute $p(\theta_t|y^i_{1:t})$ using an identification method and denote density approximation by $\tilde{p}(\theta_t|y^i_{1:t})$
    \STATE  Using $\tilde{p}(\theta_t|y^i_{1:t})$, compute parameter point estimate as ${\theta^i_{t|t}=\mathbb{E}_{\tilde{p}(\theta_t|y^i_{1:t})}[\theta_t]}$
    \ENDFOR
    \ENDFOR
    \LCOMMENT ~\begin{center} \textit{Module 3: Analysis of Bayesian identification method}  \end{center}
    \LCOMMENT ~ \textbf{Input:} Parameter sequences from Module 1, denoted by ${\{(\theta_{1:T}=\theta^i_{1:T})\}_{i=1}^M}$ and their estimates from Module 2, denoted as ${\{(\theta_{t|t}=\theta^i_{t|t})\}_{i=1,t=1}^{M,T}}$. Matrices ${\tilde{L}^\theta_t\in\mathcal{S}^q_{++}}$ and ${\tilde{P}_{t|t}^\theta\in\mathcal{S}^q_{++}}$ and tolerance level ${\epsilon\in\mathbb{R}_+^q}$ and ${\alpha\in\mathbb{R}_+^q}$
    \LCOMMENT ~ \textbf{Output:} Error analysis of identification method
    \FOR{$t=1$ to $T$}
    \STATE Compute an $M$-sample MC estimate of $\tilde{P}_{t|t}^\theta$
    \STATE Compare $\tilde{P}_{t|t}^\theta$ against $\tilde{L}_t^\theta$ 
    \STATE Compute $\{{B}^{\star,i}_{t|t}\}_{i=1}^M$ and compare against $\epsilon\in\mathbb{R}_+^q$  
    \STATE Compute an M-sample MC estimate of $\mathbb{E}_{p(y_{1:t})}[{B}^{\star}_{t|t}]$ and compare against $\alpha\in\mathbb{R}_+^q$
    \STATE Use Theorem \ref{T17} for error analysis  
     \ENDFOR
  \end{algorithmic}
\end{algorithm}
\section{Simulation example}
In this section we use a simulated system to assess the quality of a Bayesian identification  method using the procedure outlined in Algorithm \ref{alg:Algo1}. A brief introduction to the identification method considered here, is given next.
\subsection{Bayesian identification: Artificial dynamics approach}
Artificial dynamics approach (ADA) is a popular Bayesian identification method to compute $\{p(\theta_t|y_{1:t})\}_{t\in\mathbb{N}}$. In ADA, artificial dynamics is introduced to the otherwise static parameters, such that $\{\theta_t\}_{t\in\mathbb{N}}$ in (\ref{eq:E3b}) evolves according to 
\begin{equation}
\label{eq:E14}
\theta_{t+1}|\theta_t\sim \mathcal{N}(\cdot|\theta_t, Q_t^{\theta}),
\end{equation}
where ${\theta_{t+1}|\theta_t\sim \mathcal{N}(\cdot|\theta_t, Q_t^{\theta})}$ is a sequence of independent Gaussian random variable, realized independent of $\{V_t\}_{t\in\mathbb{N}}$ and $\{W_t\}_{t\in\mathbb{N}}$. By appending (\ref{eq:E3a}) and (\ref{eq:E3c}) with (\ref{eq:E14}), methods such as SMC, EKF, UKF can be used to recursively compute $\{p(\theta_t|y_{1:t})\}_{t\in\mathbb{N}}$. A detailed review on ADA can be found in  \cite{T2013} and \cite{Kantas2009}.

Even though ADA is the most widely used approach amongst the class of Bayesian identification methods, there are several standing limitations of this approach as summarized in \cite{Kantas2009} (a) the dynamics of $\{\theta_t\}_{t\in\mathbb{N}}$ in (\ref{eq:E14}) is related to the artificial noise covariance $Q_t^{\theta}$, which is often difficult to tune; and (b) adding dynamics to $\{\theta_t\}_{t\in\mathbb{N}}$ modifies the original problem, which means, it is hard to quantify the bias introduced in the estimates. 

For the former problem, the authors in see \cite{T2013} proposed an optimal rule to automatically tune $Q_t^\theta$ for all ${t\in\mathbb{N}}$; however, for the later problem, we will see how the tools developed in this paper can be used to assess the quality of ADA based Bayesian identification methods. 
\subsection{Simulation setup}
Consider the following  univariate, non stationary,  non-linear stochastic SSM (\cite{Tulsyan2013D})
\begin{subequations}
\label{eq:E13}
\begin{align}
X_{t+1}&=aX_t+\frac{X_t}{b+X_t^2}+u_t+V_t,~~V_t\sim\mathcal{N}(0,Q_t),\label{eq:E13a}\\
Y_t&=cX_t+dX_t^2+W_t,~~\quad\qquad W_t\sim\mathcal{N}(0,R_t),\label{eq:E13b}
\end{align}
\end{subequations}
where ${\theta\triangleq[a~b~c~d]}$ is a vector of unknown static model parameters. The noise covariances are constant, and selected as ${Q_t=10^{-3}}$ and ${R_t=10^{-3}}$ for all ${t\in[1,T]}$, where ${T=300}$.  $\{u_t\}_{t\in[1,T]}$ is a sequence of optimal input (see \cite{Tulsyan2013D}). For Bayesian identification of $\theta$, we define ${\{\theta_{t}=\theta_{t-1}\}_{t\in[1,T]}=\theta}$ as a stochastic process, such that ${Z_t=\{X_t,~\theta_t\}}$ is a $\mathcal{Z}$ valued extended Markov process with ${Z_0\sim\mathcal{N}(z_m,z_c)}$, where ${z_m=[1~0.7~0.6~0.5~0.4]}$, ${z_c=\diag(0.01,~0.01,~0.01,~0.01,~0.01)}$. Starting at ${t=0}$, we are interested in assessing the ADA based SMC identification method proposed in \cite{T2013}.
\subsection{Results}
Using ${M=1000}$ MC simulations, we compute the PCRLB for (\ref{eq:E13}) using Module 1 of Algorithm \ref{alg:Algo1}. Figure 1 gives the diagonal entries of ${\{\tilde{L}_t^\theta\}_{t\in[1,T]}}$. Note that amongst the four PCRLBs, the PCRLB for $b$ is the highest for all ${t\in[1,T]}$. This suggest estimation difficulties with parameter $b$. This result is not surprising, since (\ref{eq:E13}) is non-linear in parameter $b$; however, the overall decaying trend of PCRLBs in Figure 1 suggests that starting with ${\theta_0\sim p(\theta_0)}$, theoretically,  it is possible for a Bayesian identification method  to reduce the MSE associated with the parameter estimates.
\begin{figure}[h]
\centering
\label{F1}
\includegraphics[height=2.2in, width=3.5in]{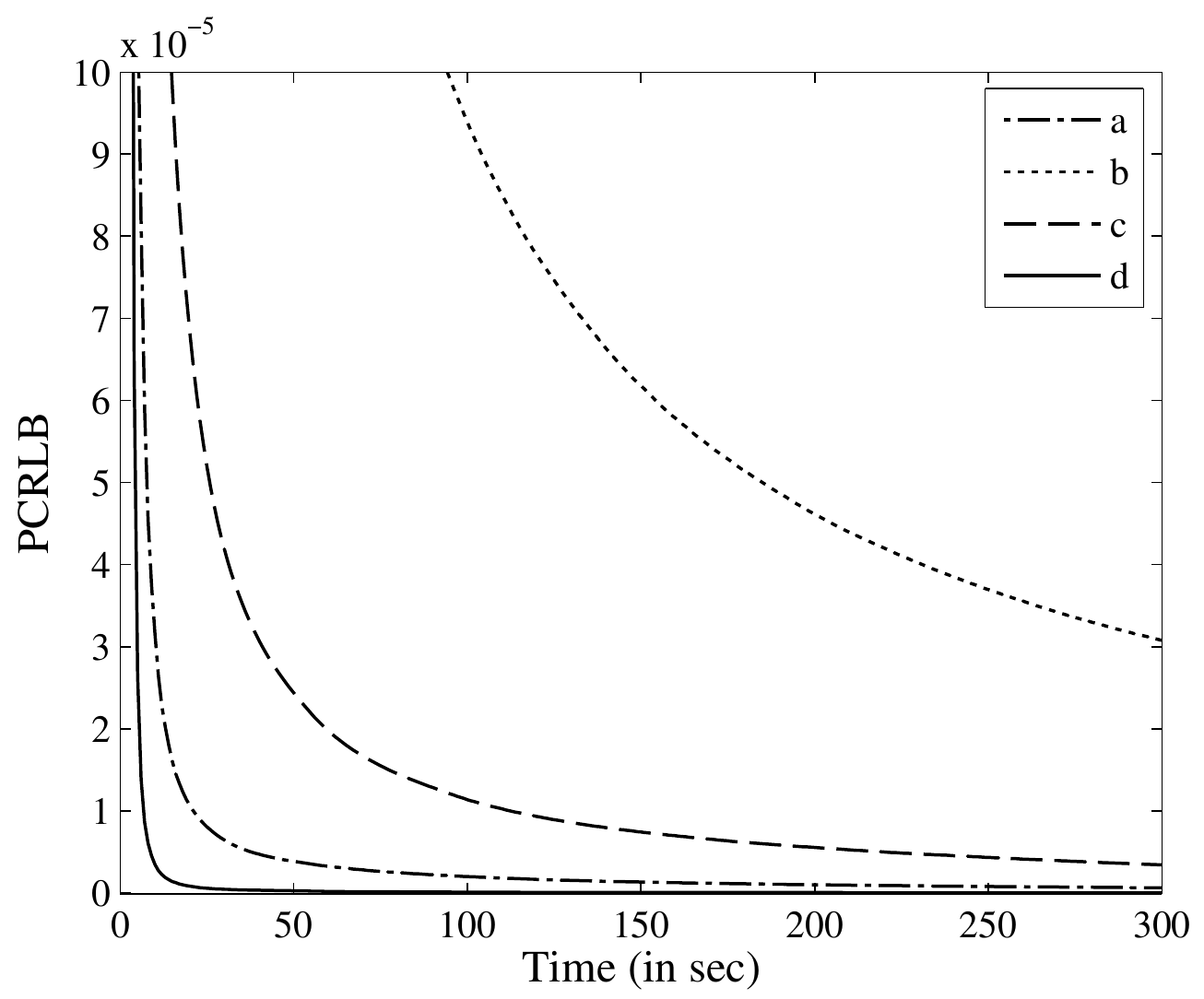} 
\caption{\small{PCRLB for parameters as a function of time. Note that the vertical axis has been appropriately scaled for clarity.}}
\end{figure}

\begin{figure}[h]
\centering
\label{figure}
\includegraphics[height=2.5in, width=3.5in]{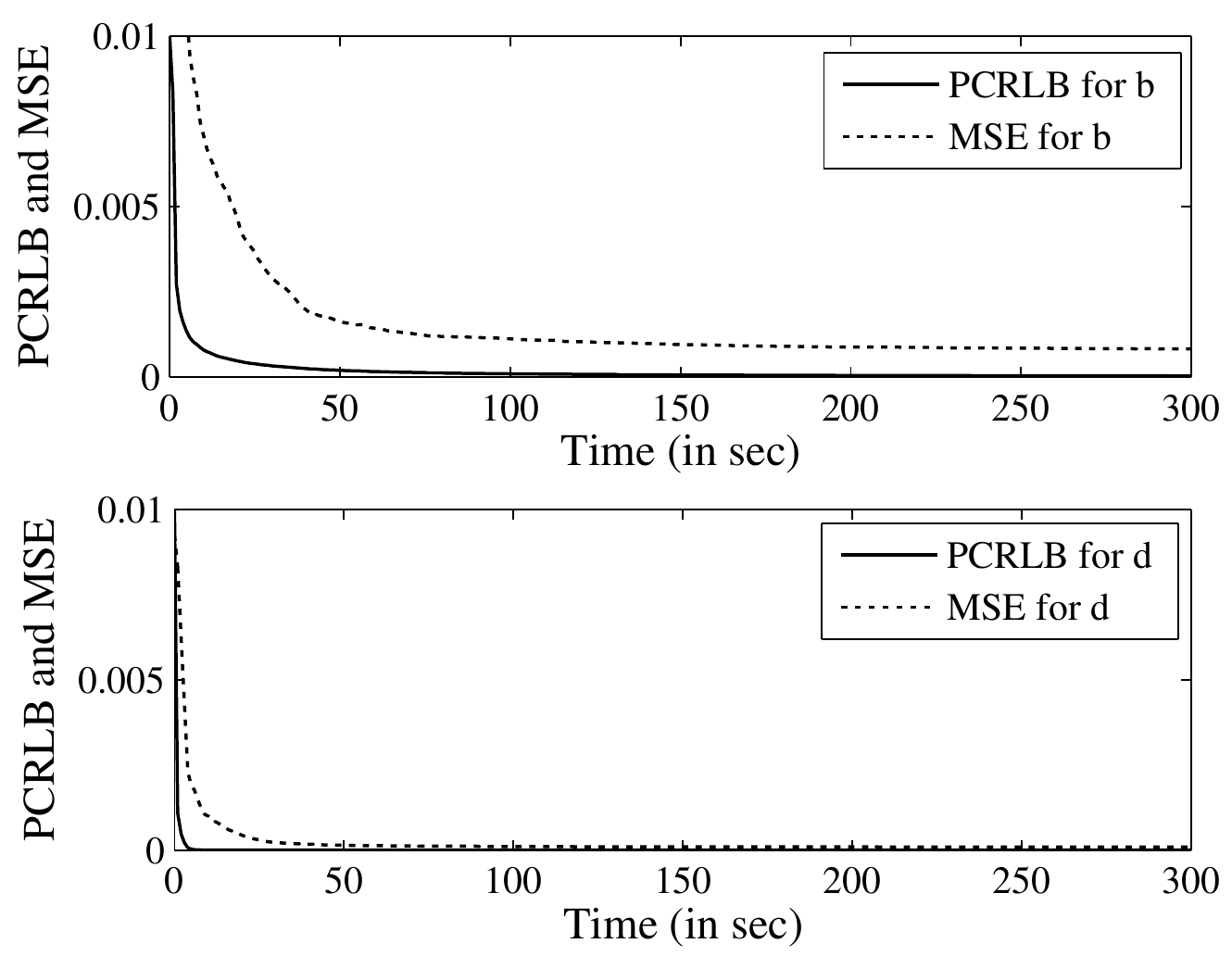} 
\caption{\small{Comparing PCRLB against the MSE computed using the ADA based SMC identification method. Graphs are shown only for parameters $b$ and $d$. Results for $a$ and $c$ are similar.}}
\end{figure}
Figure 2 compares the PCRLB against the MSE for $b$ and $d$, computed using Module 2 of Algorithm \ref{alg:Algo1}. Despite the MC approximations involved (see Remark \ref{R19}),  the MSE is greater than the PCRLB at all sampling time instants (see Figure 2). It is instructive to highlight that at $T=300$, the MSE associated with the estimation of $d$ is about $89\%$ less than that for $b$, which validates the claim made earlier about estimation difficulties  with parameter $b$. It is also important to point that for the ADA based SMC method, ${\tr[\tilde{P}^\theta_{t|t}-\tilde{L}^\theta_{t}]\neq 0}$ for all ${t\in[1,T]}$. Since the ADA based SMC method fails to satisfy the condition in Definition \ref{D3}, it is not efficient, and therefore requires error analysis. 

Figures 3 and 4 give the conditional and unconditional bias with ADA based SMC method. The results are obtained using Module 3 of Algorithm \ref{alg:Algo1}. Based on an assumed tolerance level ${\epsilon=[0.01;~0.01;~0.01;~0.01]}$ and ${\alpha=[0.001;~0.001;~0.001;~0.001]}$, in the interval ${t=[1,50]}$, less than $70\%$ of the simulations are within the specified $\epsilon$ limit (see Figure 3). Thus from Theorem \ref{T17}(b), for $t=[1,50]$, the ADA based SMC method is not even $\epsilon$-efficient, and fails to yield $\epsilon$-unbiased (except for $d$, which is $\alpha$-unconditionally unbiased, see Figure 4) or $\epsilon$-MMSE estimates. Another interesting interval is ${t=[100,T]}$; wherein, more than $70\%$ of the simulations are within the specified $\epsilon$ limit (except for parameter $b$,  where only $60\%$ of simulations are within $\epsilon$, see Figure 3). Thus from Theorem \ref{T17}(a), the ADA based SMC method is $\epsilon$-efficient for all the parameters, except for $b$, and the resulting estimates are $\epsilon$-unbiased and $\epsilon$-MMSE; whereas, for $b$, the estimates are are not MMSE, but are $\alpha$-unconditionally unbiased.

In summary, the results suggest that for model given in (\ref{eq:E13}), the ADA based SMC method at ${t=T}$ yields $\epsilon$-unbiased, $\epsilon$-MMSE estimates for all the parameters, except for parameter $b$, which is only $\alpha$-unconditionally unbiased.

\begin{figure}
\centering
\label{figure}
\includegraphics[height=2.5in, width=3.5in]{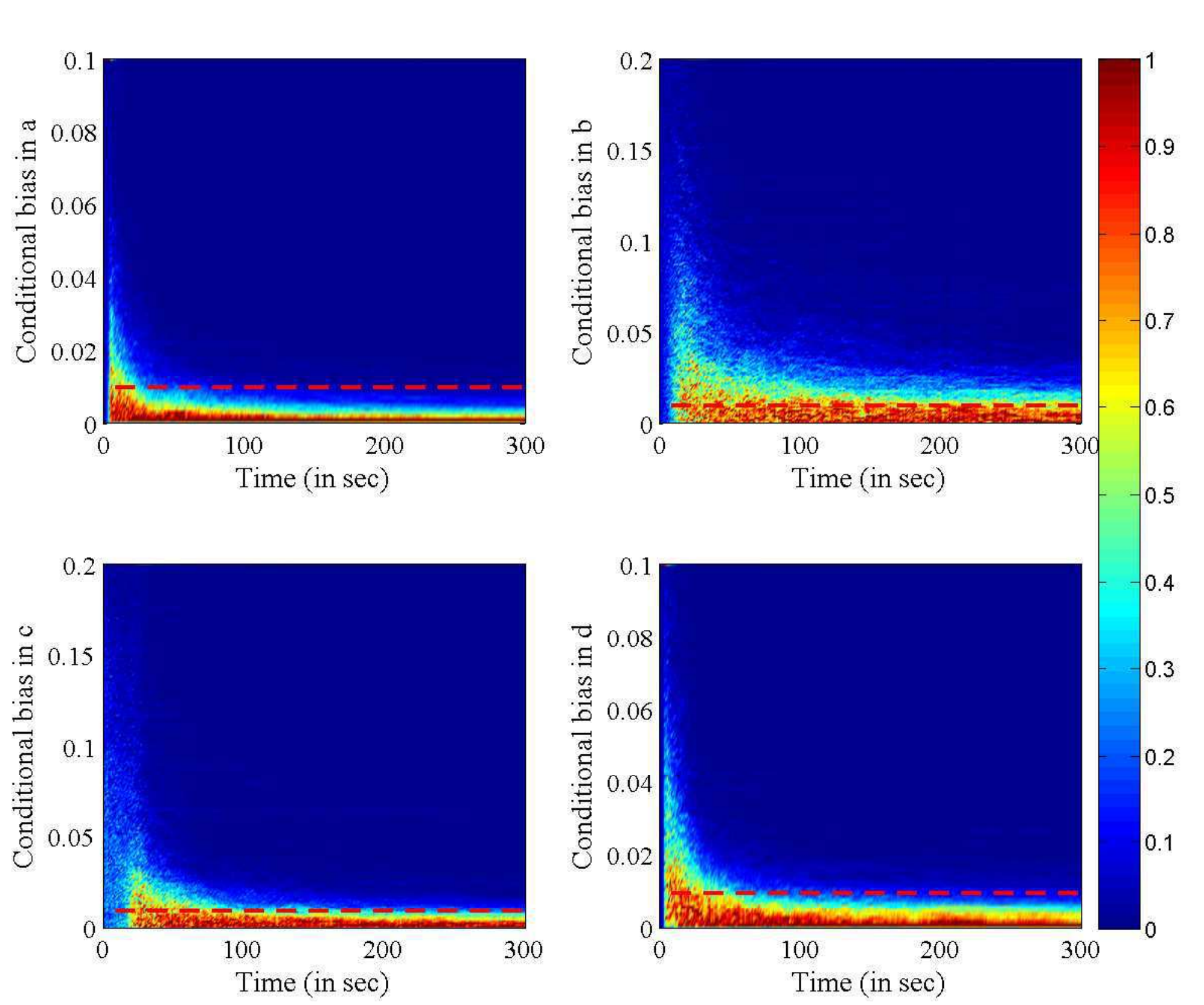} 
\caption{\small{Conditional bias in parameter estimates with ADA based SMC method. The broken red line is the $\epsilon$ value.}}
\end{figure}

\begin{figure}
\centering
\label{figure}
\includegraphics[height=2.1in, width=3.5in]{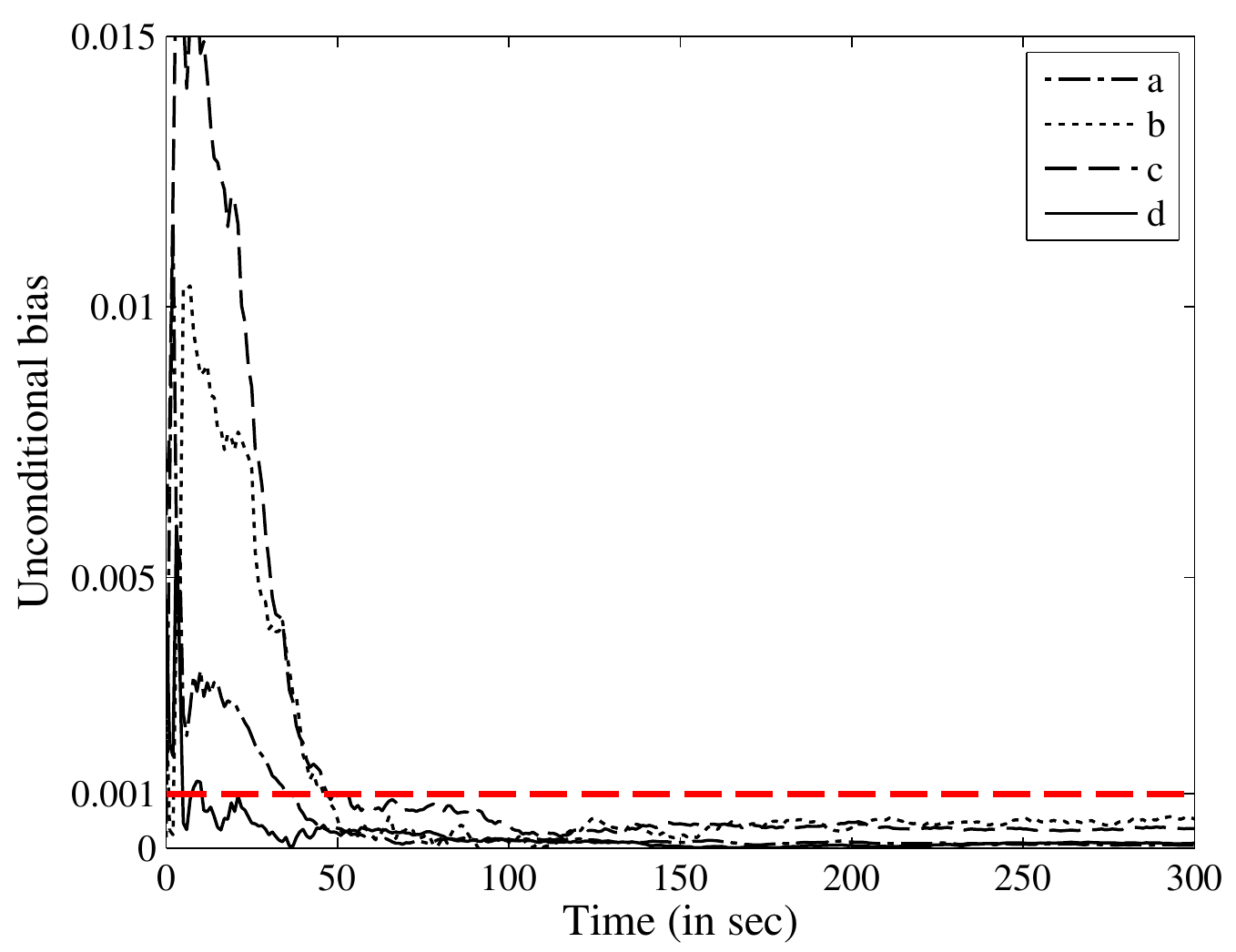} 
\caption{\small{Unconditional bias in parameter estimates with ADA based SMC method. The broken red line is the $\alpha$ value.}}
\end{figure}
\section{conclusions}
A PCRLB based approach is proposed for error analysis in Bayesian identification methods of non-linear SSMs. Using the proposed tool it was illustrated how the quality of the parameter estimates obtained using artificial dynamics approach, which is a popular Bayesian identification method can be assessed in terms of bias, MSE and efficiency.

\bibliographystyle{plain}
\bibliography{ifacconf}

\end{document}